\documentclass[final,5p,times,twocolumn,authoryear]{elsarticle}

\usepackage{amssymb}
\usepackage{stfloats}
\usepackage{lipsum}
\usepackage{xcolor} 
\usepackage{booktabs} 
\usepackage{siunitx} 
\usepackage{times}
\usepackage{bm}
\usepackage{amsmath}
\usepackage{latexsym}%
\usepackage{graphicx}%
\usepackage{amssymb}%
\usepackage{booktabs}
\usepackage[breaklinks=true,colorlinks=true]{hyperref}
\usepackage{listings}
\usepackage{algorithm}
\usepackage{algpseudocode}
\usepackage{url}
\usepackage[multiple]{footmisc}
\usepackage[many]{tcolorbox}
\usepackage{listings}
\usepackage{titlesec}
\usepackage[multiple]{footmisc}
\usepackage{listings}
\usepackage{siunitx}
\usepackage{subfigure}
\usepackage{tabularx}
\usepackage{booktabs}

\RequirePackage{color}

\lstdefinestyle{mystyle}{
    language=Python,
    backgroundcolor=\color[rgb]{0.9, 0.9, 0.9}, 
    basicstyle=\bfseries \ttfamily\footnotesize,
    breakatwhitespace=false,         
    breaklines=true,                 
    captionpos=b,                    
    keepspaces=true,                 
    numbers=left,                    
    numbersep=5pt,                  
    showspaces=false,                
    showstringspaces=false,
    showtabs=false,                  
    tabsize=2
}

\journal{Astronomy $\&$ Computing}

\begin{document}

\begin{frontmatter}

\title{\texttt{Galmoss:} A package for GPU-accelerated Galaxy Profile Fitting}

\author[1,2,3]{Mi Chen}

\author[4]{Rafael S. de Souza\corref{mycorrespondingauthor}}
\cortext[mycorrespondingauthor]{Corresponding author}
\ead{rd23aag@herts.ac.uk}

\author[1,2]{Quanfeng Xu}
\author[2,3]{Shiyin Shen}
\author[5]{Ana L. Chies-Santos}
\author[1,2]{Renhao Ye}
\author[5]{Marco A. Canossa-Gosteinski}
\author[2]{Yanping Cong}

\address[1]{School of Astronomy and Space Science, University of Chinese Academy of Sciences, 1 East Yanqi Lake Rd., Beijing 100049, P.R. China}

\address[2]{
Shanghai Astronomical Observatory, Chinese Academy of Sciences, 80 Nandan Rd., Shanghai,  200030, China
}
\address[3]{Key Lab for Astrophysics, Shanghai, 200034, China}
\address[4]{center for Astrophysics Research, University of Hertfordshire, College Lane, Hatfield, AL10 9AB, UK}
\address[5]{Instituto de Física, Universidade Federal do Rio Grande do Sul, Porto Alegre, RS, Brazil}

\begin{abstract}
We introduce \texttt{galmoss}, a \texttt{python}-based, \texttt{torch}-powered tool for two-dimensional fitting of galaxy profiles. By seamlessly enabling GPU parallelization, \texttt{galmoss} meets the high computational demands of large-scale galaxy surveys, placing galaxy profile fitting in the LSST-era. It incorporates widely used profiles such as the Sérsic, Exponential disk, Ferrer, King, Gaussian, and Moffat profiles, and allows for the easy integration of more complex models. Tested on 8,289 galaxies from the Sloan Digital Sky Survey (SDSS) g-band with a single NVIDIA A100 GPU, \texttt{galmoss} completed classical Sérsic profile fitting in about 10 minutes. Benchmark tests show that \texttt{galmoss} achieves computational speeds that are 6 $\times$ faster than those of default implementations.
\end{abstract}


\end{frontmatter}

\section{Introduction}
\label{introduction}
Galaxies, the cosmic building blocks, comprise diverse stellar components such as bulges, discs, bars, spiral arms, and nuclear star clusters. One of the main drivers of extragalactic astronomy is the study of the structural and morphological properties of galaxies from their photometric images, which has been shown to correlate with the galaxies' formation and evolutionary paths \citep{van2008dependence, conselice2014evolution, dimauro2022coincidence}.

Several approaches have been developed for galaxy structural and morphological analysis. Classical eyeball morphology classifications, such as the Hubble sequence \citep{hubble1926extragalactic}, are generally descriptive and rely on visual inspection. Non-parametric morphological analysis \citep{ferrari2015morfometryka, statemorph} employs quantitative metrics to represent attributes such as concentration, asymmetry, and smoothness \citep{conselice2003relationship}. These metrics are considered robust because they do not depend on any underlying model. In contrast, the analysis of surface brightness profiles of galaxies involves fitting the light distribution with parametric models to interpret and quantify morphology using specific model parameters \citep{nantais2013morphology, zhuang2022star}. This profile fitting method has evolved from one-dimensional \citep{de1958photoelectric, sersic1968} to two-dimensional analyses, improving accuracy by incorporating factors such as the point spread function and non-axisymmetric components \citep{andredakis1995shape, schade1995canada, byun1995two}.

The fitting of two-dimensional light profiles in galaxies has grown increasingly complex, requiring faster and more scalable algorithms. Two primary methods dominate: Markov Chain Monte Carlo (MCMC) and gradient-based approaches. MCMC-based programs, such as \texttt{profit} \citep{robotham2017profit} and \texttt{autogalaxy} \citep{pyautogalaxy}, iteratively sample to approximate the target distribution, with fitted values and uncertainties derived from the Markov chain's statistics. Conversely, gradient-based methods, utilized by tools like \texttt{galfit} \citep{peng2010galfit} and \texttt{imfit} \citep{erwin2015imfit}, offer quicker solutions by directly navigating to the loss function's local minima. However, this speed compromises the exploration of parameter space and the robustness of uncertainty estimates, a trade-off not present in MCMC-based approaches \citep{chen2016bridging}.

With the expanding volume of data from astronomical surveys such as the Sloan Digital Sky Survey \citep[SDSS,][]{york2000sloan}, Dark Energy Survey \citep[DES,][]{dark2016dark}, Chinese Survey Space Telescope \citep[CSST, ][]{zhan2011consideration}, Legacy Survey of Space and Time \citep[LSST, ][]{abell2009lsst}, and EUCLID \citep{laureijs2011euclid}, significant challenges emerge in galaxy profile fitting tasks. Both gradient-based and MCMC approaches face difficulties in processing the vast number of galaxies efficiently.

To address these challenges, deep learning strategies are increasingly being employed. For instance, \texttt{deeplegato} \citep{deeplegato} and \texttt{galnets} \citep{li2022galaxy, GaLNets} train convolutional neural networks to predict profile-fitting parameters from simulated galaxy images. These deep learning approaches establish an implicit link between galaxy images and their profile parameters, substantially accelerating the process of parameter estimation \citep[see also \texttt{gamornet},][]{ghosh2019galaxy}. However, these methods cannot be easily adapted to new data domains due to the well-known 'black box' problem \citep{castelvecchi2016can}, which means that the implicit mappings generated by the models usually lack clear interpretations.

Another method to address speed challenges involves parallel gradient-based methodology, which relies exclusively on GPU-accelerated matrix parallel computation. This is supported by popular frameworks such as \texttt{pytorch} \citep{paszke2019pytorch}, \texttt{tensorflow} \citep{abadi2016tensorflow}, and \texttt{jax} \citep{bradbury2018jax}, enabling the incorporation of automatic differentiation. Applications of this approach range from galaxy kinematic modeling \citep{bekiaris2017enhancing} to cosmological N-body simulations \citep{modi2021flowpm}. 

To speed up galaxy profile fitting, bringing it to the LSST-era and contribute to the literature on galaxy formation and evolution, we introduce \texttt{galmoss}. We aim to harness the advantages of both traditional and contemporary approaches by providing a framework that enables seamless parallelization while preserving the interpretability of more conventional methods. \texttt{galmoss} processes batches of galaxy photometric images as multidimensional arrays, facilitating efficient fitting on GPUs. This tensor-based approach permits dynamic management of data sizes during computations through the adjustment of batch sizes and the optimization of GPU utilization. For the fitting process, \texttt{galmoss} makes use of gradient descent due to its relatively low computational and memory requirements. Finally, \texttt{galmoss} provides two methods for uncertainty estimation.  In the realm of galaxy profile fitting, \texttt{astrophot} \citep{stone2023astrophot} leverages GPU acceleration and automatic differentiation in its serial optimization process while preserving physical interpretability. While our package focuses on processing numerous images of individual galaxies, \texttt{astrophot} is optimized for fitting multiple galaxies within a single, crowded image.

This paper is structured as follows: Section \ref{sec:outline} provides an overview of the \texttt{galmoss} workflow. Section \ref{sec:data} describes the data that we use to showcase the software's performance. Section \ref{sec:image} details the generation of model images, using the Sérsic profile as an example, and introduces a set of built-in profiles along with advanced applications. Section \ref{sec:fitting} discusses the fitting process post-image generation. Finally, Section \ref{sec:practice} showcases \texttt{galmoss}' performance through case studies, with concluding remarks presented in Section \ref{sec:Conclusions}.

\section{General workflow of \texttt{galmoss}}
\label{sec:outline}

\begin{figure*}[tp]
    \centering
    \includegraphics[width=0.975\linewidth]{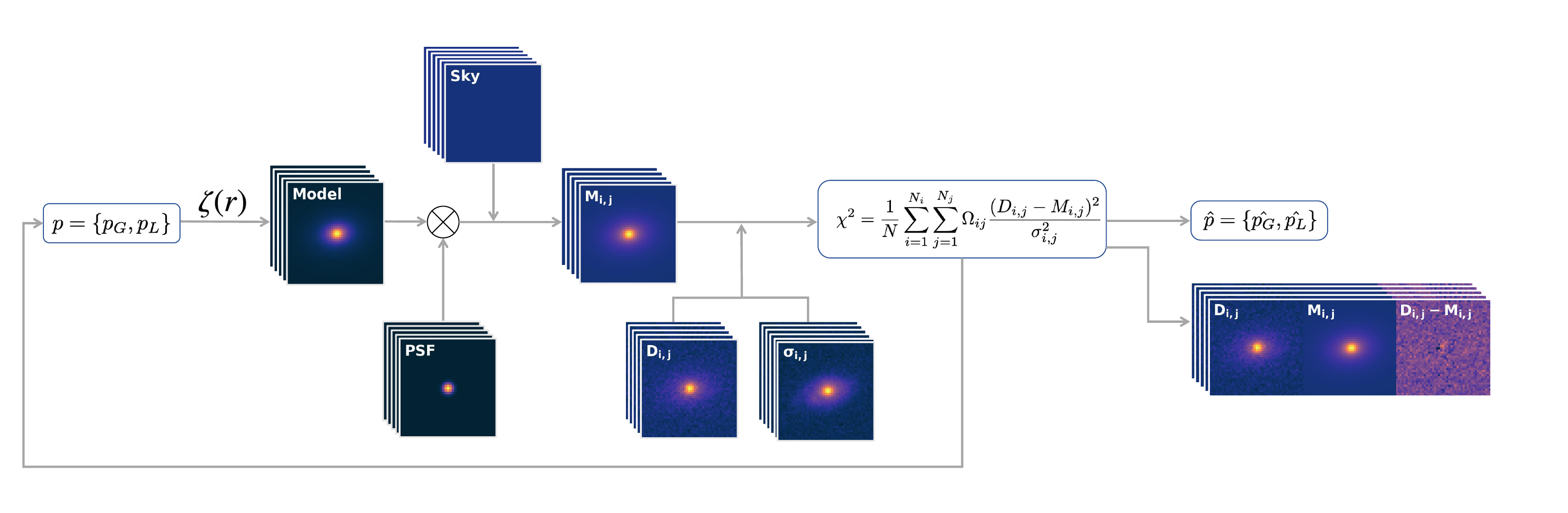}
    \caption{The \texttt{galmoss} workflow: from parameter initialization and image processing
to the fitting procedure and final data product generation, including the fitted model and
residuals, all stored in FITS format.}
    \label{fig:model}
\end{figure*}

Figure \ref{fig:model} shows the workflow of \texttt{galmoss}. It begins by reading a vector \( p  = \{p_G,p_L\}\) consisting of user-defined initial parameter values, along with the data image, sigma image (data uncertainty), mask image, and Point Spread Function (hereafter PSF) image. 
The first subset \( p_G \) describes the geometric properties, which define the central position of the ellipse's isophote, its major/minor axes, and the position angle. The second subset \( p_L \) reads the initial values for the galaxy's radial brightness profile $\mathbf{\zeta(r)}$.

With these initial values established, \texttt{galmoss} generates a series of image models following chosen profiles. These models are then convolved with the PSF of the image and augmented by adding a mean sky value to all pixels. This process results in the image model \( M_{i,j} \), which is then compared to the image data \( D_{i,j} \) and corresponding sigma images \( \sigma_{i,j} \). The fitting process employs a \(\chi^2\) loss function along with a mask \( \Omega_{i,j} \) to determine which pixels are included in the optimization. Internally, \texttt{galmoss} employs a gradient descent method embedded in \texttt{pytorch}. 

After completing the fitting process, the package saves the data products, including the image block and fitted parameter values, to the designated output path. The image block files, saved in the FITS format, comprise the original image, the best-fit model image, and the residuals. With each component stored in a given FITS Head Data Unit (HDU), in the case of fitting multiple components (e.g., bulge + disc), each component can also be accessed individually alongside other data products.

\section{Data}
\label{sec:data}

For showcasing  \texttt{galmoss} capabilities, we gathered galaxy images ($128\times128$ pixels) and PSF images ($41\times41$ pixels) from the SDSS catalog\footnote{\url{https://dr15.sdss.org/sas/dr15/eboss/photoObj/}}. Our data selection was guided by the availability of independent galaxy parameters, as provided by the MaNGA PyMorph Photometric Value Added catalog \citep[MPP-VAC-DR17,][]{dominguez2022sdss}, encompassing 10,293 galaxies from the final MaNGA release \citep{bundy2014overview}. The selected sample predominantly consists of galaxies with an average redshift \emph{z} $\sim$ 0.03 and features a relatively uniform distribution of stellar mass \citep{collaboration2022seventeenth}. The catalog offers photometric parameters obtained through two-dimensional surface profile fitting. These parameters include results from single Sérsic and combined Sérsic + Exponential models, calculated using the \texttt{pymorph} pipeline \citep{vikram2010pymorph}. \texttt{pymorph} integrates \texttt{sextractor} with \texttt{galfit}, the former package streamlines the estimation of initial parameters and the generation of stamps, and the latter aims to fit the galaxy profile.

In our selection process, we excluded galaxies located at the edges of plates where obtaining a complete $128\times128$ pixel image was not possible. We also left out galaxies that failed in the single Sérsic fitting process within the MPP-VAC-DR17 catalog. This resulted in a final sample of 8,289 galaxies. We used only g-band images to ensure a good signal-to-noise ratio throughout the sample. Following \texttt{pymorph}, we generated masked images and initial values using \texttt{sextractor} \citep{sextractor}.

\section{Image Generation and built-in Profiles}
\label{sec:image}

For a given set of galaxy or profile parameters, \texttt{galmoss} generates model images that represent the intensity or surface brightness of each pixel, based on the selected profile. \texttt{galmoss} provides six built-in radial profiles: Sérsic, Exponential disk, Ferrer, King, Gaussian, and Moffat, as well as a Flat sky model (with the option to include more complex sky models). These radial profiles are transformed into two-dimensional images by assuming circular symmetry. This process effectively projects the one-dimensional profile over a circular area in two dimensions to create the final image. To illustrate this projection process more concretely, we use the Sérsic profile as an example, given its widespread application in modeling galaxy profiles.

Furthermore, \texttt{galmoss} supports the combination and integration of different profiles. Detailed instructions on the practical use of the code can be found in the \ref{sec:bulge_disk_code}, \ref{sec: combination} and \ref{sec: used-defined}.

\subsection{Sérsic}

\label{profile}

\begin{figure*}[t]
    \centering
    \includegraphics[width=0.975\linewidth]{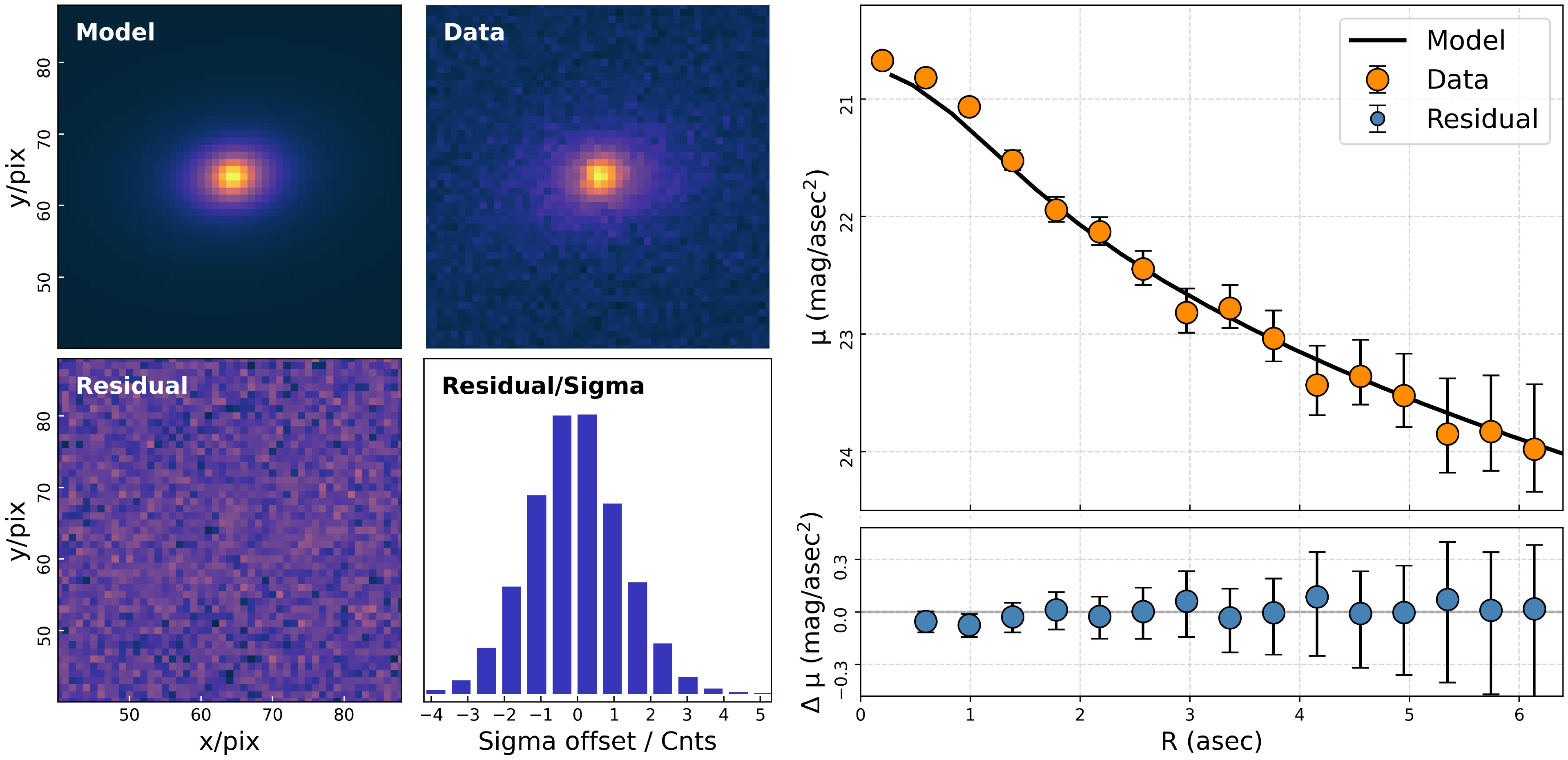}
    \caption{g-band \texttt{galmoss} fitting result of J162123.19+322056.4, showing a good fitting quality. The left four images show the model, model residuals (data-model), data, and the residual distribution. On the right, a one-dimensional projection is shown, where the upper panel represents the model distribution with a black line and the orange filled markers with error bars represent the galaxy image's flux distribution and its uncertainty. The lower panel represents the flux residual with blue filled markers with error bars.}
    \label{fig: 1-323} 
\end{figure*}

The Sérsic profile \citep{sersic1968} is commonly used to fit the surface brightness distribution of elliptical galaxies, as well as the disk and bulge components of other galaxy types.  The intensity $I(r)$ as a function of the radius $r$ is given by  

\begin{equation}\label{sersic-intensity}
I(r) = I_{\rm e} \exp \left\{
-\nu_n\left[ \left(\frac{r}{r_{\rm e}}\right)^{\frac{1}{n}} -1 \right]
\right\},
\end{equation}
where $I_{\rm e}$ denotes the surface brightness at the effective radius $r_{\rm e}$, which encompasses half of the total profile flux. The Sérsic index $n$  dictates the profile's curvature, and the parameter $\nu_n$ is computed numerically from the inverse cumulative distribution function of the gamma distribution.  The versatility of the Sérsic profile lies in its ability to model a wide range of galaxy morphologies by adjusting the Sérsic index $n$. For example, setting $n=4$ transforms it into the de Vaucouleurs profile, known as the $r^{1/4}$ profile. An exponential disk profile is obtained with $n=1$, and $n=0.5$ results in a Gaussian profile, enabling users to explore these diverse profiles by simply varying $n$.

In a one-dimensional context, the radius $r$  refers to the distance from the galaxy center. In two dimensions, given a galaxy profile center $(x_{\rm{c}}, y_{\rm{c}})$, the position angle of the ellipse profile $\theta$, and the axis ratio $q$, the circular radius $r$ is defined by \citep[see e.g.][]{robotham2017profit}:

\begin{align}
\label{r}
r &= \left[r_{\rm{maj}}^{(B + 2)} + \left(\frac{r_{\rm{min}}}{q}\right)^{(B + 2)}\right]^{\frac{1}{B + 2}}, 
\end{align}
where
\begin{align}
r_{\rm{maj}} &= \left| \cos(\theta)(x-x_{\rm{c}})\right| + \left| \sin(\theta)(y-y_{\rm{c}})\right|, \label{rmaj} \\
r_{\rm{min}} &= \left| -\sin(\theta)(x-x_{\rm{c}})\right| + \left| \cos(\theta)(y-y_{\rm{c}})\right|. \label{rmin}
\end{align}

The boxiness parameter $B$ introduces flexibility in the shape of the ellipse \citep{athanassoula1990shape}, allowing for more general elliptical profiles. $B = 0$ corresponds to a standard ellipse, while positive values result in more box-like shapes and negative values in more disc-like shapes. This flexibility is particularly useful in galaxy modeling, where a range of elliptical shapes can represent the diverse morphologies observed in galaxies.

Rather than directly using the surface brightness $I_{\rm e}$ in the Sérsic, magnitude $m$ and its corresponding zero point magnitude $m_{0}$ is used to specify the intensity level. This approach is consistent with methodologies employed by widely-used galaxy fitting tools such as \texttt{galfit} \citep{peng2010galfit} and \texttt{profit} \citep{robotham2017profit}, facilitating easier comparison with observational data. Both quantities are related as follow:

\begin{equation}
\label{sersicLtot} 
\quad I_{\rm e} =  \frac{r_{\rm box}\nu_n^{2n} 10^{-0.4(m - m_{0})} }{2 \pi q r_{\rm e}^2 n \Gamma(2n) \exp(\nu_n)},
\end{equation}
where
\begin{equation}
r_{\rm box} = \frac{\pi(2+B)}{2\beta\left(\frac{1}{2+B}, \frac{1}{2+B}\right)}.
\label{rbox}
\end{equation}

In this formulation, $r_{\rm box}$ serves as a geometric correction factor in $I_{\rm e}$ to account for deviations from a perfect ellipse, influenced by the level of diskiness or boxiness. Specifically, when $B$ = 0, indicating a perfect ellipse,  $r_{\text{box}}$ = 1, implying no geometric correction. The beta function $\beta(a, b)$, is calculated using the relationship  $\beta(a, b) = \frac{\Gamma(a)\Gamma(b)}{\Gamma(a+b)}$, where $\Gamma$ denotes the Gamma function.

Figure \ref{fig: 1-323} presents the fitting results using the Sérsic profile for galaxy J162123.19+322056.4, which is one of the galaxies in our catalog. The quality of this model is evidenced by the residuals, highlighting the precision of the fit. For those interested in the implementation details, the code used for this analysis is provided in \ref{sec:example_single_sersic}.

\subsection{Other available profiles}
\label{other_profile}
In this section, we briefly discuss the built-in profiles other than Sérsic. All these profiles but the Flat sky profile have circular radius $r$ that follows Equation \ref{r} to Equation \ref{rmin}.

\subsubsection{Exponential Disk Profile}
\label{sec:expdisk}
In an exponential disk profile, the intensity is defined as follows:

\begin{equation}\label{expdisk_I} 
    I(r) = I_0 \exp \left( -\frac{r}{r_{\rm_{s}}} \right),
\end{equation}

where $I_0$ is the brightness of the profile's surface in the center (at radial distance $r=0$), and $r_{\rm_{s}}$ is the disk scale-length. 

The relation among the magnitude ($m,m_0)$, $r_{\rm_{s}}$ and $I_0$ is given by

\begin{equation}\label{expdiskI0}
    I_0 = \frac{r_{\mathrm{box}} 10^{-0.4(m-m_0)}}{2 \pi q r_{\mathrm{s}}^2 },
\end{equation}

where $q$ is the axis ratio, and $r_{\rm{box}}$ is as defined in equation \ref{rbox}.

\subsubsection{Modified Ferrer Profile}

The modified Ferrer profile, which is characterised by a nearly flat core and a rapidly truncated shape on the periphery, was originally proposed by \cite{ferrers} and later modified by \cite{modifiedferrers}. The intensity is defined as follows:

\begin{equation}\label{ferrer_I}
    I(r) = I_0 \left[ 1 - \left( \frac{r}{r_{\mathrm{out}}} \right)^{2-b} \right]^a,
\end{equation}

where \(I_0\) is also the central surface brightness parameter, \(r_{\mathrm{out}}\) is the outer truncation radius, and \(b\) and \(a\) are parameters governing the slopes of the truncation and core, respectively. 

The relation between the magnitude (\(m\), \(m_0\)) and the profile model parameters is given by

\begin{equation}\label{ferrerI0}
    I_0 = \frac{r_{\mathrm{box}} 10^{-0.4(m-m_0)}}{\pi q r_{\mathrm{out}}^2 a \beta \left(a, 1+\frac{2}{2-b}\right)},
\end{equation}
where $q$ is also the axis ratio and $\beta$ is the beta function defined in equation \ref{rbox}. The modified Ferrer profile is particularly suitable to model the bar structure in galaxies \citep{BarUse1, BarUse2}.

\subsubsection{Empirical (Modified) King Profile}

Since its initial illustration by \cite{king1962structure}, the empirical King profile has been extensively used to fit both galactic and extragalactic globular clusters \citep[BAOlab\footnote{\url{https://github.com/soerenslarsen/baolab}} presented by][]{larsen2014baolab, chies2007high, bonatto2008structural, tripathi2023photometric}. The intensity of this profile is defined as follows:

\begin{equation}
\label{BE_S}
\begin{split}
    I(r) = & I_0 \left[ 1 -  \frac{1}{ \left( 1 + \left(\frac{r_{\rm t}}{r_{\rm c}}\right)^2 \right) ^{1/ \alpha}} \right]^{-\alpha} \\
           & \times \left[ \frac{1}{\left(1 + \left(\frac{r_{\rm t}}{r_{\rm c}}\right)^2\right)^{1/ \alpha}} -  \frac{1}{\left(1 + \left(\frac{r}{r_{\rm c}}\right)^2\right)^{1/ \alpha}} \right]^{\alpha}.
\end{split}
\end{equation}

Here, $I_0$ is also the central surface brightness parameter. The core radius, $r_{\rm c}$, signifies the scale at which the density starts to deviate from uniformity, while $r_{\rm t}$, the truncation radius, marks the boundary of the cluster. The global power-law factor, $\alpha$, dictates the rate at which the density declines with distance from the center. The concentration of stars in globular clusters can be defined using the parameters $r_{\rm{t}}$ and $r_{\rm{c}}$, denoted by $c = \log \left( \frac{r_{\rm{t}}}{r_{\rm{c}}} \right)$.

\begin{figure*}[tp]
    \centering
    \includegraphics[width=0.975\linewidth]{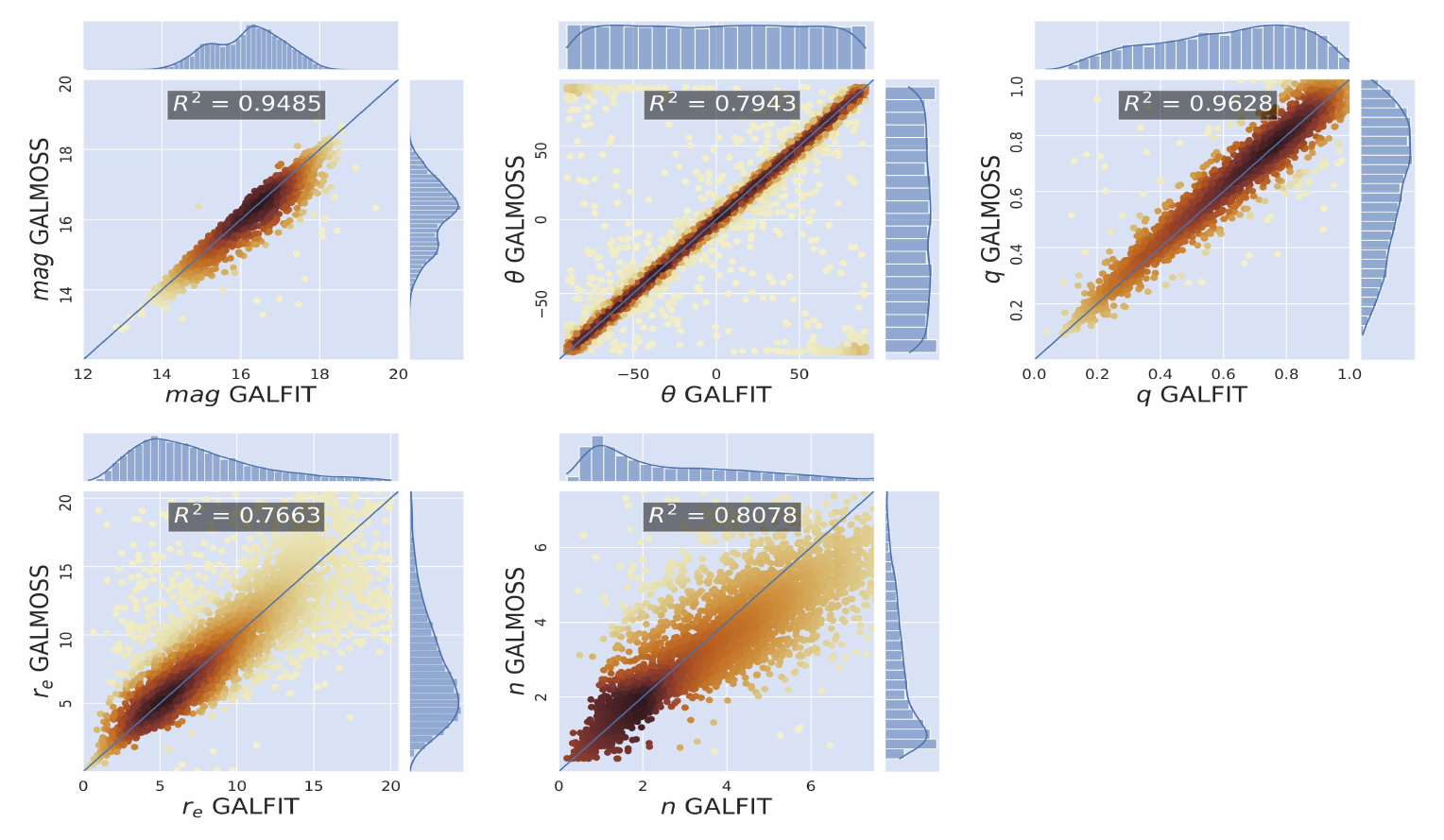}
    \caption{Panels comparing the measured Sérsic profile parameters for the selected $\sim$ 8,000 SDSS galaxies, along with $R^2$ estimation. The panels each display galaxy magnitude ($m$), position angle ($\theta$), axis ratio ($q$), effective radius ($r_e$) and Sérsic index ($n$). The transparency of the dots illustrates their concentration. The blue line shows where \texttt{galfit} result equals \texttt{galmoss} result. The histogram on top and right illustrates the distribution of measured values for \texttt{galmoss} and \texttt{galfit}, respectively. The comparison shows proper consistency in all panels.}
    \label{fig:result}
\end{figure*}

\subsubsection{Gaussian Profile}
\label{sec:Gaussian}

In a Gaussian profile, the intensity is defined as follows:

\begin{equation}\label{gaussion_I} 
    I(r) = I_0 \exp \left( -\frac{r^2}{2\sigma^2} \right),
\end{equation}

where $I_0$ is also the central surface brightness parameter, and $\sigma$ is the radial dispersion. In the \texttt{galmoss} implementation, the Full Width at Half Maximum (FWHM =  2.354$\sigma$) is used instead.  The classical application of the Gaussian profile includes modelling a simple PSF and point sources.

\subsubsection{Moffat Profile}

The Moffat profile \citep{moffat},  commonly used for modelling a realistic telescope PSF, defines its intensity as follows:
\begin{equation}\label{moffat_I}
    I(r) = I_0 \left[ 1 + \left(\frac{r}{r_{\rm d}}\right)^2 \right ] ^{-n},
\end{equation}
where 
\begin{equation}\label{moffat_rd}
    r_{\rm d} = \frac{\rm FWHM}{2\sqrt{2^{\frac{1}{n}} -1}}.
\end{equation}
and  $I_0$ also is the central density. The concentration index, $n$, dictates whether the distribution is more Lorentzian-like ($n$ =1) or Gaussian-like  (\( n \to \infty \)).

 When considering  the ellipse with  axis ratio $q$ and boxiness parameter ($r_{\rm{box}}$ in equation \ref{rbox}), the relationship between the observed magnitude $m$ and profile model parameter is 
\begin{equation}\label{moffatI0}
    I_0 = \frac{r_{\rm box} (n-1) 10^{-0.4(m-m_{\rm{0}})}}{\pi q r_{\rm d}^2}.
\end{equation}

Similar to the Gaussian profile, the Moffat profile can also be used to model point sources.

\subsubsection{Flat Sky}
\label{profile: sky}

Unlike other radial light profiles, in \texttt{galmoss}, the sky profile is controlled by the sky mean value $I_{\rm{sky}}$ across all pixels without a radial matrix:

\begin{equation}\label{sky_I}
    I = I_{\rm{sky}}.
\end{equation}

In \texttt{galmoss}, the only parameter needed in sky Profile is
$I_{\rm{sky}}$. If a sky profile varies with position (radius), it can be included as a user-defined profile.

\section{Image Fitting and Evaluation}
\label{sec:fitting}
Following the generation of model images, \texttt{galmoss} implements an extended \(\chi 2\)  likelihood and gradient descent optimization, leveraging \texttt{python} and \texttt{pytorch} functionalities. In addition, uncertainty estimation is available.

\subsection{Maximum Likelihood Estimation}
\label{subsec:Likelihood}

The \texttt{galmoss} package employs a $\chi^2$ likelihood:
\begin{align}
\label{chinu}
&\ln \mathcal{L} = - \frac{k}{2}\ln{2} - \ln\Gamma(k/2)+
 \left(\frac{k}{2} - 1 \right)\ln{\chi^2} -  \dfrac{\chi^2}{2},\\
 &\chi ^2 = \sum_{i=1}^{N_i} \sum_{j=1}^{N_j} \Omega _{i.j} \frac{(D_{i,j} - M_{i,j})^2}{\sigma^2_{i,j}}, 
\end{align}
where $M_{i,j}$ represents the model at pixel $i,j$, and $\sigma_{i,j}$ represents the uncertainty of $D_{i,j}$. In the $\chi^2$ distribution, the degrees of freedom, $k$, is the difference between the number of pixels with galaxy flux that are not included in the mask and the number of model fit parameters.

Internally \texttt{galmoss} adopts gradient descent for parallel fitting. Though slower in convergence than other traditional methods (e.g. Levenberg-Marquardt) it has lower computational and memory demands, enabling efficient parallel fitting on GPUs.

\subsection{Confidence Intervals}

\begin{figure}[t]
    \centering
    \includegraphics[width=0.975\linewidth]{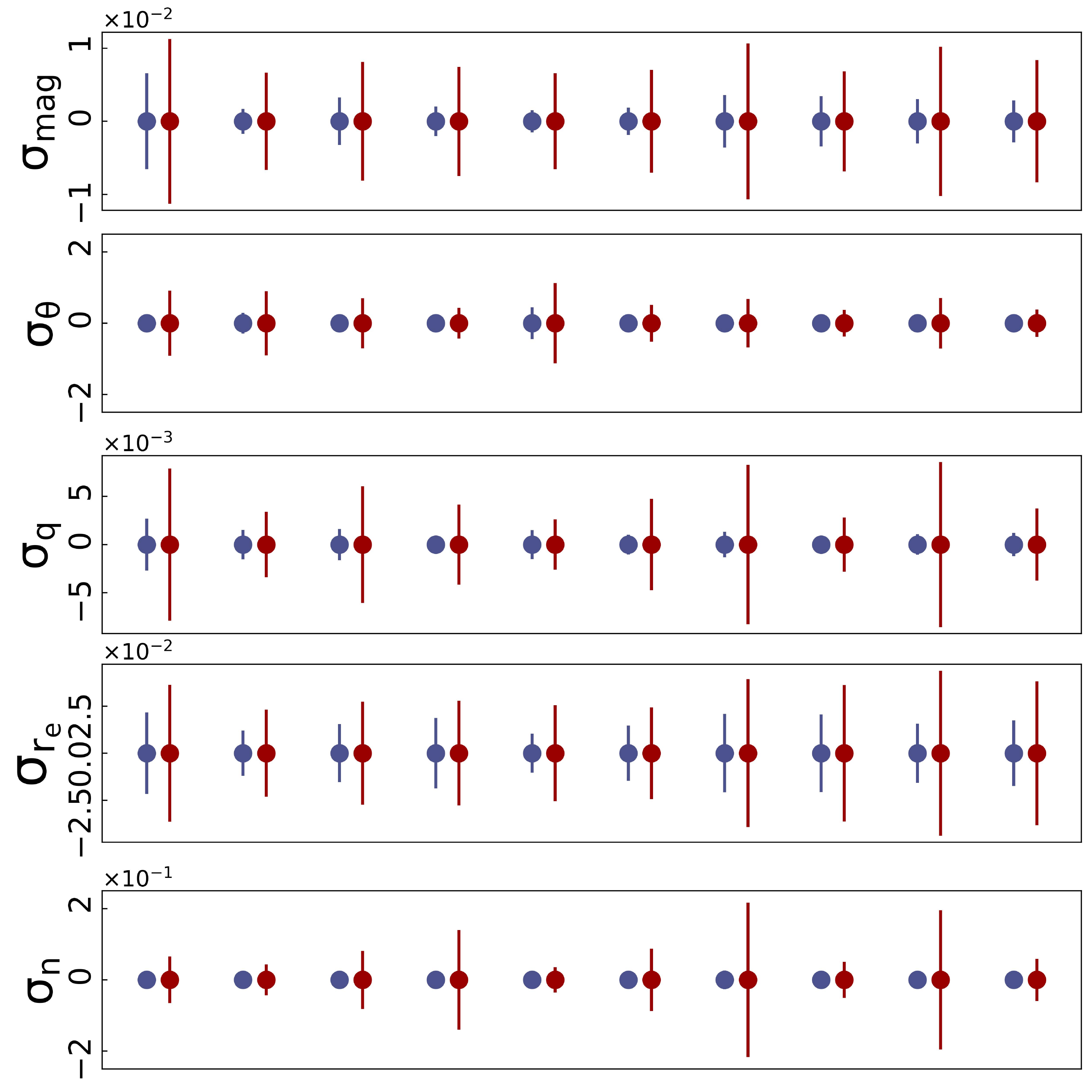}
    \caption{A illustrative comparison of uncertainty estimations for fitted galaxy parameters using two methods: covariance matrix (depicted with blue error bars) and bootstrapping (depicted with red error bars), for a random selection of galaxies within our sample. Typically, the uncertainties derived from bootstrapping are observed to be larger than those obtained from the covariance matrix.}
    \label{fig:uncertainty} 
\end{figure}

\texttt{Galmoss} employs two strategies to estimate uncertainties in galaxy images, primarily using a covariance matrix based on Gaussian distribution assumptions \citep{uncertainty}. The parameter uncertainties are calculated from the covariance matrix's diagonal, derived using the Jacobian matrix \citep[e.g.,][]{lm} for computational efficiency. This method approximates uncertainties as $\sigma_{\rm{p}} = \sqrt{\mathrm{diag} \left[J^{T}WJ\right]^{-1}}$, where $J$ is the Jacobian matrix, and $W$ is diagonal with $W_{i,j} = 1/\sigma^2_{i,j}$, offering a 1-$\sigma$ confidence level for the parameters.  The second methodology for uncertainty estimation is bootstrapping, which involves resampling with replacement each galaxy images multiple times (typically 100 iterations) and refitting them. The uncertainty is given by $\sigma^2 = \frac{1}{M} \sum_{i=1}^{M} \left [ m_i - \bar{m} \right ]^2$, where $\bar{m}$ is the mean value of the estimated parameter. 
Figure \ref{fig:uncertainty} illustrates a comparison of uncertainty estimation methods for a random subset of galaxies. We display the uncertainties determined through bootstrap analysis (red error bars) and those derived from the covariance matrix (blue error bars). Generally, bootstrap uncertainties encompass a broader range, suggesting a more conservative estimate.


\begin{figure}[t]
    \centering
    \includegraphics[width=0.975\linewidth]{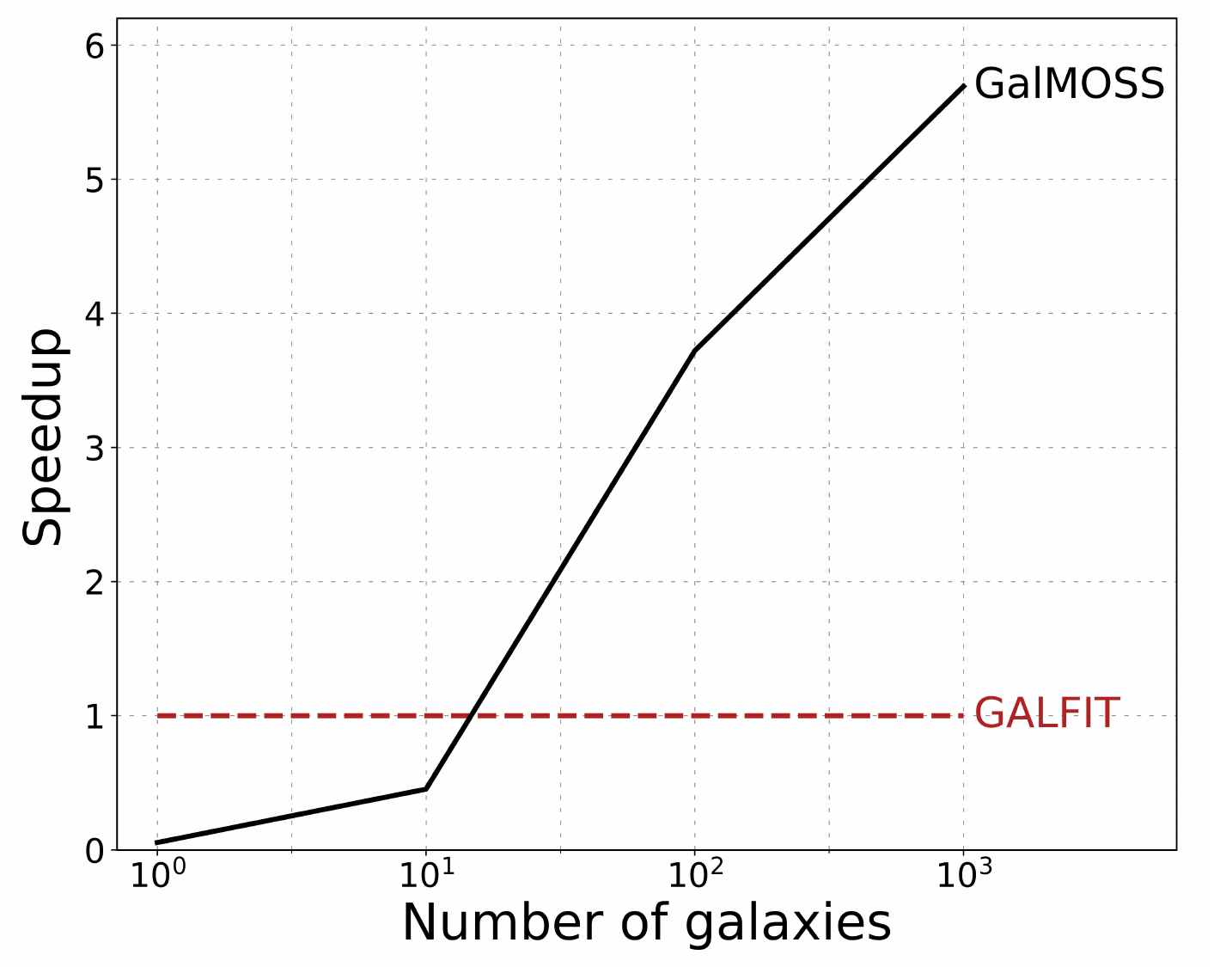}
    \caption{A speed-up performance evaluation between \texttt{galmoss} and \texttt{galfit}. Each line
represents the speedup time normalized by the running time of \texttt{galfit}. The results show that \texttt{galmoss}’s speed surpasses \texttt{galfit} when the batch size
slightly exceeds 10 and about 6 $\times$ faster when the batch size is 1,000. The benchmark was executed
in a machine with the following specifications: CPU – 2.2GHz Intel Xeon Silver 4210; GPU – NVIDIA A100 (80GB); OS – Ubuntu Linux 18.04 64 bits; RAM – 3.9 TB.}
    \label{speedup} 
\end{figure}

\section{Results}
\label{sec:practice}

To evaluate the performance of \texttt{galmoss} against established methods, we performed a validation test by fitting single galaxy profiles from the same dataset and comparing the structural parameters, such as the Sérsic index and effective radius, with those obtained using \texttt{galfit}, as catalogued in the MPP-VAC-DR17. This comparison acts as a fundamental 'sanity check' to verify the reliability of \texttt{galmoss}'s fitting results. Figure \ref{fig:result} shows overall consistency across key galaxy parameters—magnitude ($m$), position angle ($\theta$), axis ratio ($q$), effective radius ($r_{\rm{e}}$), and Sérsic index ($n$)—as quantified by the $R^2$ coefficient of determination.

The panels for magnitude, position angle, and axis ratio show a strong linear correlation, indicative of a high degree of alignment with \texttt{galfit} results. Specifically, in the position angle panel, objects that deviate significantly from the identity line are those fitted with large axis ratios ($q \sim 1$). This suggests that, in these cases, the light distributions are relatively insensitive to variations in the position angle ($\theta$).

The parameters $r_{\rm{e}}$ and $n$, known to be challenging to fit accurately \citep{trujillo2001estimation}, demonstrate a linear relationship, albeit with greater dispersion compared to other parameters. Notably, in the Sérsic index panel, \texttt{galmoss} values are generally lower than those derived from \texttt{galfit} at higher values of $n$. This discrepancy could be attributed to variations in image quality and the characteristics of the optimization algorithms used in \texttt{pytorch}. For an in-depth analysis of this observed bias, we direct the reader to Section \ref{bias}.

In addition to the aforementioned test, we conducted an independent speed comparison using \texttt{galfit} and \texttt{galmoss} using the same dataset and initial values. The fitting of 8,289 galaxies with \texttt{galmoss} completed in roughly 10 minutes—six times faster than with serial \texttt{galfit}.  Figure \ref{speedup} demonstrates this efficiency gain across different batch sizes, showing \texttt{galmoss}' speed advantage becoming more pronounced as batch size increases, up to the limits of GPU capacity.

\section{Conclusions}
\label{sec:Conclusions}

In this study, we introduce \texttt{galmoss}, an open-source software package specifically designed for fitting galaxy light profiles on large datasets, ideally suited for the LSST era. Built on the torch framework, \texttt{galmoss} provides efficient 2D surface brightness profile fitting for batches of galaxy images, with the added benefit of GPU acceleration. It features a user-friendly interface that allows for the easy definition of parameters and their ranges. Moreover, \texttt{galmoss} is capable of efficiently quantifying uncertainty through both covariance matrix and bootstrap methods. The package also supports the integration of new profiles, making it an adaptable and versatile tool for statistical analysis in galaxy structure studies. 

Benchmark tests reveal that \texttt{galmoss} can achieve speedup gains of up to 6 times compared to \texttt{galfit}, depending on the sample size. One novel aspect of \texttt{galmoss} is its use of native parallelization to process multiple observational fields of single galaxies as elements of a single multidimensional tensor, thereby enhancing its speed through efficient \texttt{PyTorch} GPU-accelerated matrix calculations and memory usage \citep{paszke2019pytorch}. This approach contrasts with \texttt{astrophot}, which focuses on larger fields with multiple galaxies or joint multi-band fitting. A current limitation of \texttt{galmoss} is the requirement to manually select profile models and initial parameters, along with its sub-optimal GPU utilization for small galaxy batches—areas for improvement in future versions.

\texttt{galmoss} is freely
available at GitHub \footnote{\url{https://github.com/Chenmi0619/GALMoss/}}, Zenodo \footnote{\url{https://doi.org/10.5281/zenodo.10654784}}
and listed in the Python
Package Index \footnote{\url{https://pypi.org/project/galmoss/}}. The readthedocs \footnote{\url{https://galmoss.readthedocs.io/en/latest/}} contains example usages, along with an overview of the package.

\section*{Acknowledgments}
We thank Gongyu Chen for helping with the logic and language flow of the paper. We thank MengTing Shen for helping with the preparation of SDSS dataset. We thank Zhu Chen and Ziqi Ma for their help in learning the usage of \texttt{sextractor}. We thank Qiqi Wu and Shuairu Zhu for their help in \texttt{galmoss} application. 

ACS acknowledges funding from the brazilian agencies \textit {Conselho Nacional de Desenvolvimento Cient\'ifico e Tecnol\'ogico} (CNPq) and the Rio Grande do Sul Research Foundation (FAPERGS) through grants CNPq-11153/2018-6 and FAPERGS/CAPES 19/2551-0000696-9.  SS thanks research grants from the China Manned Space Project with NO. CMS-CSST-2021-A07, the National Key R\&D Program of China (No. 2022YFF0503402, 2019YFA0405501,), National Natural Science Foundation of China (No. 12073059 \& 12141302 ) and Shanghai Academic/Technology Research Leader (22XD1404200).

Funding for the Sloan Digital Sky Survey IV has been provided by the Alfred P. Sloan Foundation, the U.S. Department of Energy Office of Science, and the Participating Institutions. SDSS acknowledges support and resources from the Center for High-Performance Computing at the University of Utah. The SDSS web site is www.sdss4.org.

\appendix

\section{How to use \texttt{galmoss}: single Sérsic case}
\label{sec:example_single_sersic}

Here, we demonstrate how to fit a single Sérsic profile to SDSS image data using the \texttt{galmoss} package.

First, we need to load the necessary packages.
\begin{lstlisting}
import Galmoss as gm
\end{lstlisting}

Next, we need to define the parameter objects and associate them with profile instances. The initial estimates of the galaxy parameters are provided by \texttt{sextractor}. Notably, we do not include the boxiness parameter in this simple example, despite its availability within the \texttt{galmoss} framework.

\begin{lstlisting}
# define parameter objects and profile

sersic = gm.lp.Sersic(
    cen_x=gm.p(65.43),
    cen_y=gm.p(64.95),
    pa=gm.p(-81.06, angle=True), 
    axis_r=gm.p(0.64),
    eff_r=gm.p(7.58, pix_scale=0.396),
    ser_n=gm.p(1.53, log=True),
    mag=gm.p(17.68, M0=22.5)
)

\end{lstlisting}

The comprehensive dataset object can be formulated utilising the image sets (galaxy image, sigma image, PSF image, mask image) together with the chosen profiles.
\begin{lstlisting}
dataset = gm.Dataset(
      galaxy_index="J162123.19+322056.4",
      image_path="./J162123.19+322056.4_image.fits",
      sigma_path="./J162123.19+322056.4_sigma.fits",
      psf_path="./J162123.19+322056.4_psf.fits",
      mask_path="./J162123.19+322056.4_mask.fits"
      mask_index=2,
      img_block_path="./test_repo",
      result_path="./test_repo"    
)

dataset.define_profiles(sersic=sersic)

\end{lstlisting}

After initializing the hyperparameter during the fitting process, fitting could start. Subsequently, we run the uncertainty estimation process.

\begin{lstlisting}
fitting = gm.Fitting(dataset=dataset, 
                     batch_size=1, 
                     iteration=1000)
fitting.fit()
fitting.uncertainty(method="covar_mat")

\end{lstlisting}

When the fitting process is completed, the fitted results and the img\_blocks are saved in corresponding path.

\section{How to use \texttt{galmoss}: bulge+disk case}
\label{sec:bulge_disk_code}

\begin{figure*}[t]
    \centering
    \includegraphics[width=0.975\linewidth]{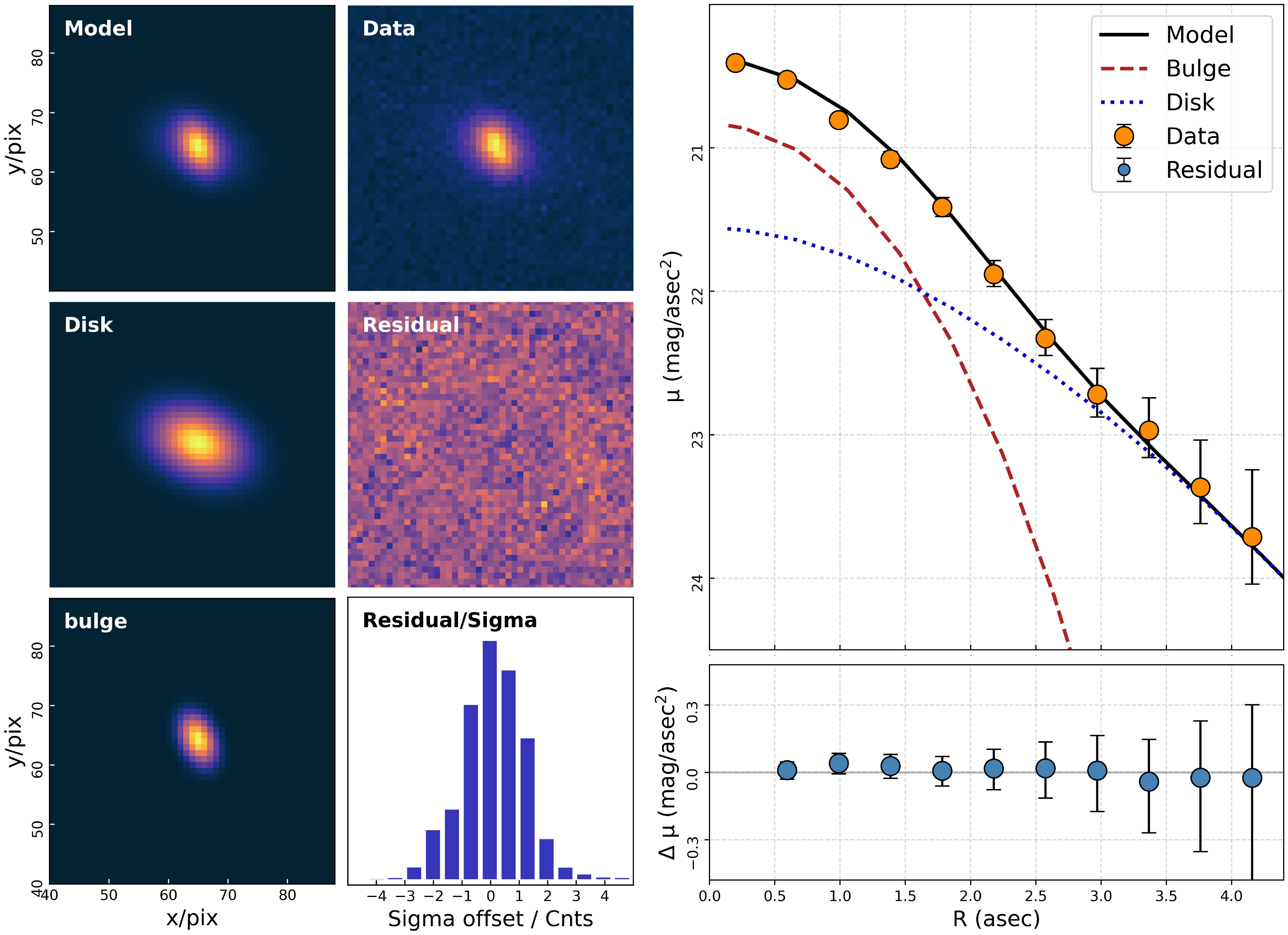}
    \caption{g-band \texttt{galmoss} decomposition result of J100247.00+042559.8, showing a good fitting quality. The left six images show the total model, disk model, bulge model, data, model residuals (data-model), and the residual distribution. On the right, a one-dimensional projection is shown. The upper panel represents the model distribution with a black line, and the orange dots with error bars represent the galaxy image's flux distribution and its uncertainty. The red dashed line and blue dotted line represent the distribution of the bulge component and the disk component, respectively. The lower panel represents the flux residual with blue dots with error bars.}
    \label{fig: 1-78702} 
\end{figure*}

Here, we demonstrate how to use a combination of two Sérsic profiles to make disk and bulge decomposition on SDSS image data using the \texttt{galmoss} package.

\begin{lstlisting}
import Galmoss as gm
\end{lstlisting}

Upon importing the package, the subsequent step entails defining parameter objects. To ensure that the center parameter within both profiles remains the same, it suffices to specify the center parameter once and subsequently incorporate it into various profiles.
\begin{lstlisting}
xcen = gm.p(65.97)
ycen = gm.p(65.30)
\end{lstlisting}

For a quick start, we let the disk and bulge profile share the initial value from the \texttt{sextractor}, with an initial Sérsic index of 1 for the bulge component and 4 for the disk component.  
\begin{lstlisting}
bulge = gm.lp.Sersic(
    cen_x=xcen,
    cen_y=ycen,
    pa=gm.p(58.70, angle=True), 
    axis_r=gm.p(0.75),
    eff_r=gm.p(4.09, pix_scale=0.396),
    ser_n=gm.p(4),
    mag=gm.p(17.97, M0=22.5)
)

disk = gm.lp.Sersic(
    cen_x=xcen,
    cen_y=ycen,
    pa=gm.p(ini_value=58.70, angle=True), 
    axis_r=gm.p(0.75),
    eff_r=gm.p(ini_value=4.09, pix_scale=0.396),
    ser_n=gm.p(ini_value=1),
    mag=gm.p(ini_value=17.97, M0=22.5)
)
                
\end{lstlisting}
Compared to the single profile case, we only need to change the code snippet of profile definition. We choose to use bootstrap to calculate the uncertainty here.

\begin{lstlisting}
dataset = gm.Data_Box(
      galaxy_index="J100247.00+042559.8"],
      image_path="./J100247.00+042559.8_image.fits",
      sigma_path="./J100247.00+042559.8_sigma.fits",
      psf_path="./J100247.00+042559.8_psf.fits",
      mask_path="./J100247.00+042559.8_mask.fits"
      img_block_path="./test_repo",
      result_path="./test_repo"
)
              
dataset.define_profiles(bulge=bulge, disk=disk)
fitting = gm.Fitting(dataset=dataset, 
                     batch_size=1, 
                     iteration=1000)
fitting.fit()
fitting.uncertainty(method="bstrap")
\end{lstlisting}

Figure \ref{fig: 1-78702} show the decomposition result, with the residual demonstrating a high quality of fitting.

\section{Example code for definition of a set of galaxy profiles}
\label{sec: combination}

\begin{figure}
    \centering
    \subfigure[The Model image without convolving a PSF image.]{
        \includegraphics[width=0.47\columnwidth]{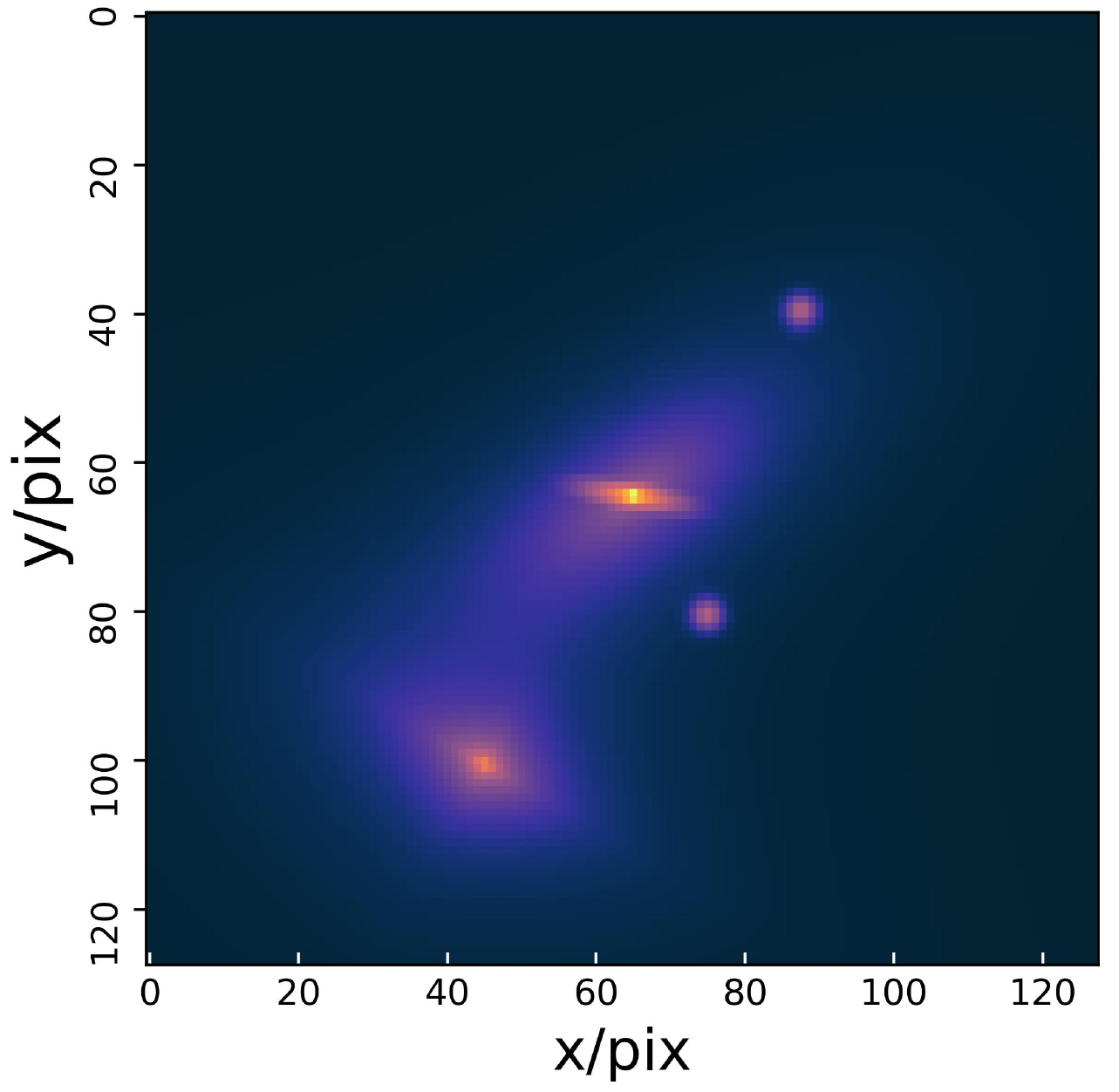}
        \label{fig:com_n}
    }
    \subfigure[The Model image with a PSF image convolved.]{
        \includegraphics[width=0.47\columnwidth]{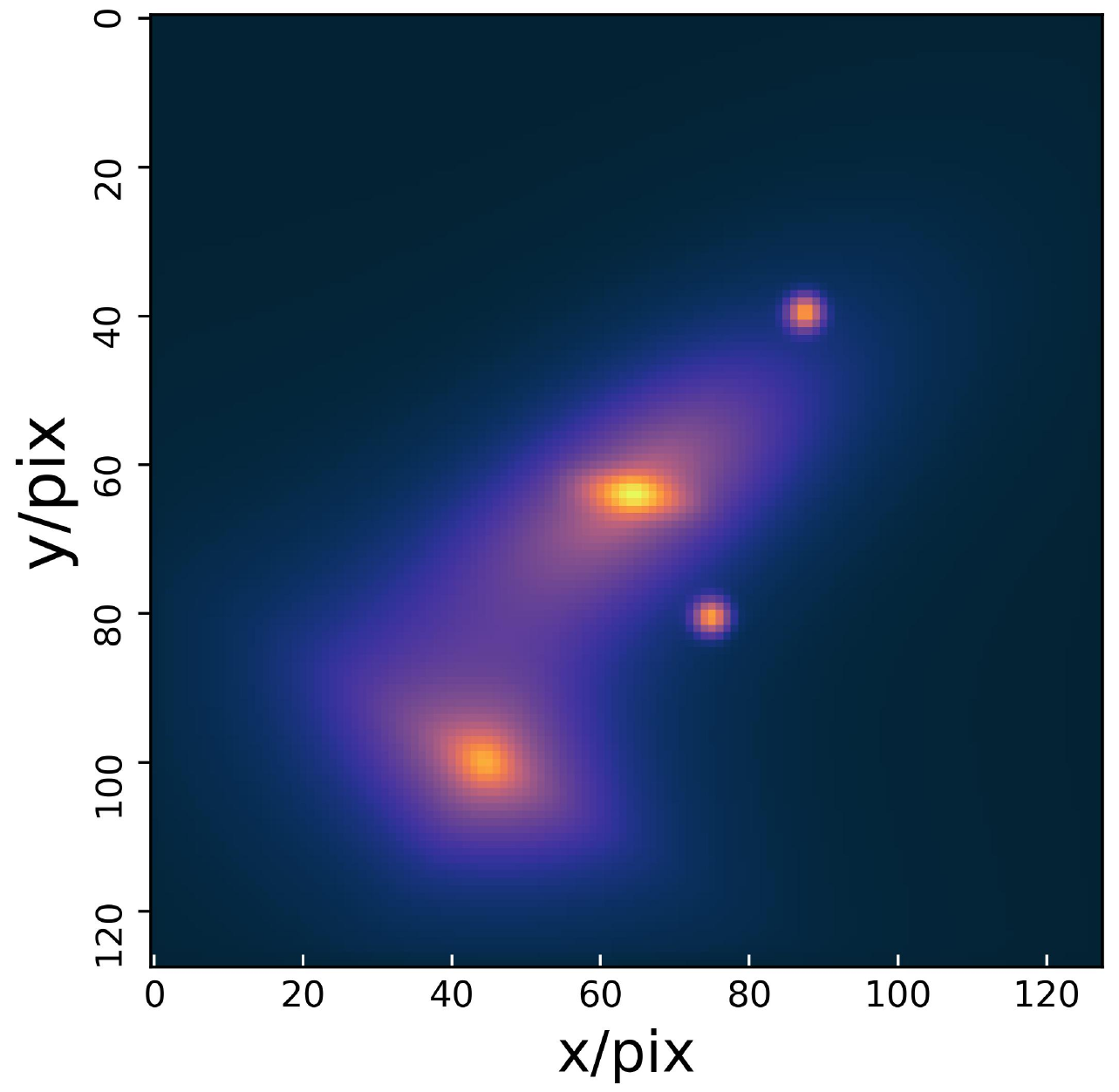}
        \label{fig:com}
    }
    \caption{The images of the model are provided for an initial review. The model consists of a central galaxy that has a disk, bulge, and bar, each defined by two Sérsic profiles and a Ferrer profile, respectively. Additionally, there is a side galaxy defined by a Sérsic profile, two point sources defined by Gaussian profile.  }
    \label{fig:combination}
\end{figure}

Here, we provide an example that shows how to define a combination of profiles and extract the model image from the image function for an initial review. Figure \ref{fig:combination} shows a model image featuring a central galaxy with a disk, bulge, and bar, each defined by two Sérsic profiles and a Ferrer profile, respectively. Additionally, there is a side galaxy defined by a Sérsic profile, two point sources defined by Moffat profiles, and an open cluster defined by a King profile. All of these are convolved with a PSF image produced by a Gaussian profile.

\begin{lstlisting}
import numpy as np
import Galmoss as gm
import torch_optimizer as optim
\end{lstlisting}

Upon importing the package, the first sub-model represents the central galaxy, comprising a disk, bulge, and bar.
\begin{lstlisting}
# The galaxy center
xcen = gm.p(65.)
ycen = gm.p(64.)

# Define parameters and profiles
Bisk = gm.lp.Sersic(
    cen_x=xcen,
    cen_y=ycen,
    pa=gm.p(55., angle=True), 
    axis_r=gm.p(0.44),
    eff_r=gm.p(13, pix_scale=0.396),
    ser_n=gm.p(1),
    mag=gm.p(17, M0=22.5),
    box=gm.p(0.1)
)

Bulge = gm.lp.Sersic(
     cen_x=xcen,
     cen_y=ycen,
     pa=gm.p(0., angle=True), 
     axis_r=gm.p(1),
     eff_r=gm.p(6, pix_scale=0.396),
     ser_n=gm.p(6),
     mag=gm.p(21.5, M0=22.5)
)

Bar = gm.lp.King(
   cen_x=xcen,
   cen_y=ycen,
   pa=gm.p(-40., angle=True), 
   axis_r=gm.p(0.2),
   mag=gm.p(21.5, M0=22.5),
   trunc_r=gm.p(5, pixScale=0.396),
   trunc_a=gm.p(0.5),
   trunc_b=gm.p(1.9),
   box=gm.p(0.1)
)  
                              
\end{lstlisting}

The second sub-model is a side galaxy.

\begin{lstlisting}
Sersic_s = gm.lp.Sersic(
        cen_x=gm.p(45.43),
        cen_y=gm.p(100.95),
        pa=gm.p(-60., angle=True), 
        axis_r=gm.p(0.7),
        eff_r=gm.p(13, pix_scale=0.396),
        ser_n=gm.p(1.6),
        mag=gm.p(17.5, M0=22.5),
        box=gm.p(-0.6)
)
\end{lstlisting}

The third sub-models are two point sources. They share the dispersion $\sigma$ = \SI{1.515}{\arcsecond} with the idealizsed PSF image, both of which are produced by a Gaussian profile.

\begin{lstlisting}
# Two point-souce
P1 = gm.lp.Gaussian(
  cen_x= gm.p(75), 
  cen_y= gm.p(80), 
  pa=gm.p(0., angle=True), 
  axis_r=gm.p(1), 
  inten=gm.p(0.1), 
  fwhm=gm.p(1.2,pix_scale=0.396),
)
    
P2 = gm.lp.Gaussian(
  cen_x=gm.p(88.),
  cen_y=gm.p(40.),
  pa=gm.p(0., angle=True), 
  axis_r=gm.p(1),
  inten=gm.p(0.1),
  fwhm=gm.p(1.2, pix_scale=0.396),
)

# The idealized PSF
PSF = gm.lp.Gaussian(
   cen_x=gm.p(20),
   cen_y=gm.p(20),
   pa=gm.p(0., angle=True), 
   axis_r=gm.p(1),
   inten=gm.p(1/(2*np.pi*((0.6/0.396)**2))),
   fwhm=gm.p(1.2, pix_scale=0.396),
)


\end{lstlisting}

The total model image without convolution is the sum of three sub-components (see Figure C.\ref{fig:com_n}). To extract the model images, it is necessary to first define a grid, as grid generation occurs automatically only during the fit process within the fitting object.

\begin{lstlisting}
x = torch.linspace(0.5, 128-0.5, 128)
y = torch.linspace(128-0.5, 0.5, 128)
grid = torch.meshgrid(y, x, indexing='ij')     
\end{lstlisting}

\begin{lstlisting}
bulge = Bulge.image_via_grid_from(grid, mode="initial_model")
disk = Disk.image_via_grid_from(grid, mode="initial_model")
bar = Bar.image_via_grid_from(grid, mode="initial_model")
sersic_s = Sersic_s.image_via_grid_from(grid, mode="initial_model")
p1 = P1.image_via_grid_from(grid, mode="initial_model")
p2 = P2.image_via_grid_from(grid, mode="initial_model")
result = bulge + disk + bar + sersic_s + p1 + p2

\end{lstlisting}

To mimic the effect of seeing, an idealized PSF image will be generated from a relatively small grid and then convolved with the model images (except for the point sources).   The result is shown in Figure C.\ref{fig:com}.

\begin{lstlisting}
#Grid generation
x_psf = torch.linspace(0.5, 40-0.5, 40)
y_psf = torch.linspace(40-0.5, 0.5, 40)  
grid_psf = torch.meshgrid(x_psf, y_psf, indexing='ij')   

# PSF image
psf = PSF.image_via_grid_from(grid_psf, mode="initial_model")

#Convolving
result_conv = torch.squeeze(torch.nn.functional.conv2d((bulge+disk+bar+sersic_s).reshape(1,1, 128,128), psf.reshape(1, 1, 40, 40), padding="same", groups=1)) + p1 + p2
\end{lstlisting}

\section{User-defined profile}
\label{sec: used-defined}

In Section \ref{profile: sky}, we introduced the sky profile. Here, we show how to integrate a more realistic sky profile as an example of a user-defined profile. In this new sky profile, the sky intensity is defined as follows:

\begin{equation}\label{new_sky_I}
    I = I_0 + k_{\rm_{x}}(x - x_{\rm{0}}) + k_{\rm_{y}}(y - y_{\rm{0}}),
\end{equation}
where $(x_{\rm{0}}, y_{\rm{0}})$ is the geometric center of the image, $I_0$ is the sky value at $(x_{\rm{0}}, y_{\rm{0}})$, and $(k_{\rm_{x}}, k_{\rm_{y}})$ describes the variation of the sky value along the x and y axes.

To integrate this new profile, we need to define a new class, which we have named $NewSky$. This class should include a function that guides the generation of the image model from the given parameters. Here are some caveats.  Firstly, every profile class should inherit from the basic class \textit{LightProfile} to access general light profile functions. Next, the parameters should be loaded into the profile class in the \textit{\_\_init\_\_} function, and set as attributions (e.g., self.sky\_0 = sky\_0).

\begin{lstlisting}
import galmoss as gm

class NewSky(gm.LightProfile):
    def __init__(self, sky_0, grad_x, grad_y):
        super().__init__()
        self.psf = False
        self.sky_0 = sky_0
        self.grad_x = grad_x
        self.grad_y = grad_y
    
    def image_via_grid_from(self,
                            grid, 
                            mode="updating_model"):
        return (self.sky_0.value(mode)
                + self.grad_x.value(mode)
                * (grid[1]-(grid[0].shape[1] + 1)/2) 
                + self.grad_y.value(mode)
                * (grid[0]-(grid[0].shape[0] + 1)/2))

\end{lstlisting}

The equation for the profile is defined within the \textit{$image\_ via\_ grid\_ from$} function. Parameter values are extracted after the mode value, which has a default value of \textit{updating\_model}. This mode calls for values that are continuously updated throughout the fitting process and have already been broadcast to a suitable shape for multi-dimensional matrix calculations.

A newly defined profile can be used as follows:

\begin{lstlisting}
sky = NewSky(sky_0=gm.p(0.3),
             grad_x=gm.p(2),
             grad_y=gm.p(3))
\end{lstlisting}

\section{The fitted bias in high Sérsic index end}
\label{bias}
As shown in Figure \ref{fig:result}, we observe a general bias in the high Sérsic index ($n$) end in the comparison results. Galaxies with large $n$ values (e.g., $n>6$) in the \texttt{galfit} results mostly exhibit lower values (e.g., $n < 6$) in the \texttt{galmoss} results.

\begin{figure}
    \centering
    \includegraphics[width=0.975\linewidth]{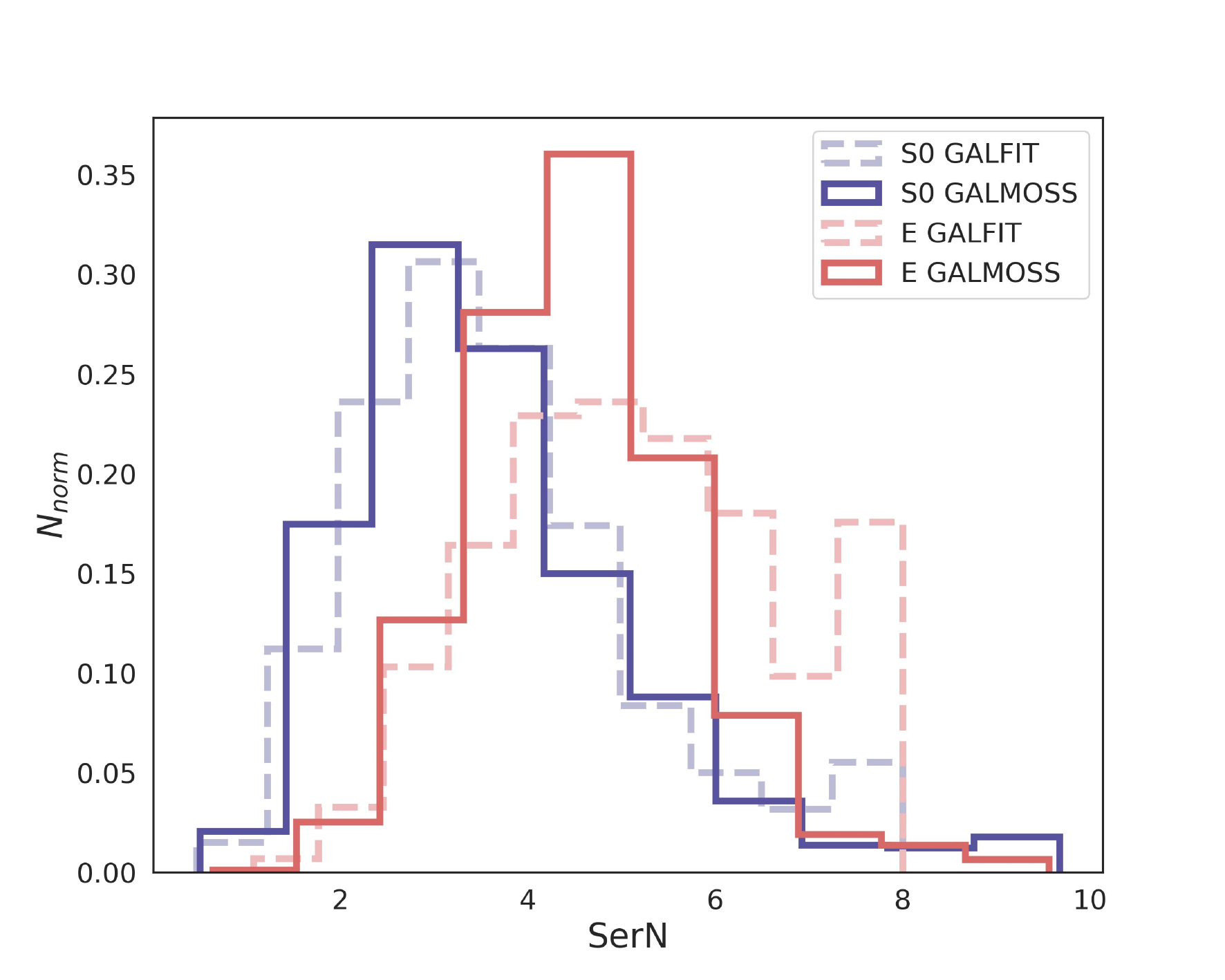}
    \caption{The histogram shows the distribution of $n$ values in E and S0 galaxies. Solid lines represent \texttt{galmoss} results, and dashed lines are for \texttt{galfit}. Purple and pink correspond to S0 and E, respectively. It is evident from the inconsistency between the values fitted by \texttt{galmoss} and \texttt{galfit} that Elliptical galaxies contribute most to the inconsistency in the high end of $n$ fitted results.}
    \label{fig:hist} 
\end{figure}

To further investigate the type of high $n$ galaxies, we use the morphological classification afforded by catalog MDLM-VAC-DR17. We identify 8,289 galaxies into 2,849 ETGs and 5,440 LTGs, and further identify ETGs into 1,891 Elliptical galaxies, 793 S0 galaxies and 165 undefined galaxies. We find that ETGs account for 34.37 $\%$ of the total galaxy samples and represent 88.78 $\%$ in the biased region. To investigate further, we plot a histogram for the $n$ value in E and S0 (Figure \ref{fig:hist}). In the plot, the histogram almost coincides between \texttt{galmoss} fitted results and \texttt{galfit} fitted results in S0 galaxies. However, the distribution of the histogram is very different between the \texttt{galmoss} fitted results and the \texttt{galfit} fitted results in elliptical galaxies. This result reveals that Elliptical galaxies contribute the most to the inconsistency in fitted $n$ values.

There are several reasons for the observed result. One of them is the limitation of observation. Most elliptical galaxies have a steep central core and a flat outer wing. Due to the poor signal-to-noise ratio in the flat wing, the $\chi ^2$ is insensitive to the change of $n$, which acts as a small gradient during the optimization. Another possible reason is related to the properties of the optimizer. Most prevalent optimizers in \texttt{pytorch} use self-adaptive learning rates which reduce the learning rate under low gradient conditions to fasten convergence. As a result, the sensitivity of $n$ is further reduced at the high-value end compared to \texttt{galfit}, which adds to the difference in results between the two software in this case.

\bibliographystyle{elsarticle-harv} 
\bibliography{ref}

\end{document}